\documentclass[]{JHEP3}
\usepackage{epsfig,multicol,bbm}

\title{Effects of the galactic magnetic field upon \\
large scale anisotropies of extragalactic Cosmic Rays}

\author{D. Harari, S. Mollerach and E. Roulet\\
CONICET, Centro At\'omico Bariloche,
Bustillo 9500, 8400 Bariloche, Argentina \\ 
E-mail: \email{harari@cab.cnea.gov.ar,mollerach@gmail.com,eroulet@gmail.com}}

\abstract{
The large scale pattern in the arrival directions 
of extragalactic cosmic rays that reach the Earth is different 
from that of the flux arriving to the halo of the Galaxy as a result of the
propagation through the galactic magnetic field. Two different effects are 
relevant in this process: deflections of trajectories and (de)acceleration 
by the electric field component due to the galactic rotation.
The deflection of the cosmic ray 
trajectories makes the flux intensity arriving to the halo from some 
direction to appear reaching the Earth from another direction. 
This applies to any intrinsic anisotropy in the 
extragalactic distribution or, even in the absence of intrinsic anisotropies, 
to the dipolar Compton-Getting anisotropy induced when the observer is moving 
with respect to the cosmic rays rest frame. For an
observer moving with the solar system, cosmic rays traveling through 
far away regions of the Galaxy also experience an electric force 
coming from the relative motion (due to the rotation of the Galaxy) of the
local system in which the field can be considered as being purely magnetic. 
This produces small changes in the particles
momentum that can originate large scale anisotropies even for an isotropic 
extragalactic flux.}


\begin{document}

\section{Introduction}

The Compton-Getting effect is a dipolar anisotropy due to the Doppler
shift of the particle spectrum when the observer is moving with
respect to the cosmic rays (CR) rest frame \cite{CG}. Interpretation of
large scale anisotropy observations requires a knowledge of the
contribution due to the Compton-Getting effect. The amplitude $\Delta$ of 
the resulting dipole depends on the velocity of the observer $v$ and on
the cosmic ray spectral index $\gamma$, through $\Delta = (2+\gamma) v/c$.
The effect has been observed at the solar daily frequency by several 
experiments in the 0.1 - 100 TeV energy range. At this frequency the 
motion of the Earth around the Sun with a velocity of $\sim$ 29.8 km/s
gives rise to a dipolar type anisotropy of amplitude 
$\Delta \sim 5 \times 10^{-4}$ and maximum at a phase $\phi \sim 6$ hs in 
local solar time,
in good agreement with the observations \cite{cu86,am04}. 
At the sidereal daily frequency the computation of the expected 
Compton-Getting dipolar anisotropy is more uncertain
as the rest frame of the local cosmic rays is not known. 
At relatively low energies 
the cosmic rays are mostly of Galactic origin and are trapped for a long 
time in the galactic magnetic field. The first harmonic modulation in right 
ascension (which is equivalent to the sidereal time) 
for energies up to a few hundred TeV, measured by different
experiments to be $\sim$ few $\times 10^{-4}$, indicates that cosmic
rays corotate with the local environment around the Galactic center 
\cite{am06,gu07,ab09,ag09}.
If the cosmic rays rest frame were instead linked to an inertial frame
fixed to the center of the Galaxy, the solar system velocity of
$\sim$ 220 km/s around the galactic center would lead to a much larger 
($\Delta \sim$ few $\times 10^{-3}$) dipolar anisotropy.

It has been pointed out by Kachelriess and Serpico \cite{KS}
that the Compton-Getting effect could also induce an observable anisotropy
of extragalactic cosmic rays. At the highest energies, cosmic rays
are not expected to be confined by the Galactic magnetic field and 
it is believed 
that they are mostly of extragalactic origin. The energy at which the 
transition from the galactic to the extragalactic
components occurs is still an open question. According to some models, the
ankle in the spectrum at $\sim$ 4 EeV (where 1 EeV $\equiv 10^{18}$ eV)
is due to a transition from a steep 
galactic spectrum to a flatter extragalactic one \cite{HSW}, while
another model explains the ankle as a result of the dip produced by 
$e^{+}e^{-}$ pair production by extragalactic protons interacting with
the cosmic microwave background (CMB) \cite{BGG}. In the later case, 
the galactic to extragalactic transition would take place at energies 
below 1 EeV.

The large scale distribution observed at Earth 
of extragalactic cosmic rays should be 
affected by the Compton-Getting effect. For energies below the Greisen
Zatzepin Kuz'min (GZK) threshold of $\sim 60$ EeV
they can come from cosmological distances. A reasonable assumption 
is that the cosmic ray flux at these energies should look isotropic in the 
CMB rest frame. 
The observed dipole in the CMB temperature allows to 
deduce the solar system velocity with respect to the CMB rest frame,
that is $v = 371 \pm 1$ km/s in the $(b,l) = (48.4^{\circ},264.4^{\circ})$
  direction in galactic coordinates \cite{fi96}.
This motion would induce a dipole with amplitude $\Delta \simeq
0.006$ (for spectral index $\gamma \simeq 3$) due to the Compton-Getting effect.
The presence of magnetic fields will however modify the structure of the 
dipolar anisotropy because the deflections of the particle trajectories lead
to a non-trivial mapping between the directions of incidence at the halo and 
the arrival directions at Earth. 
The galactic magnetic field, with its regular
component, is expected to lead to the largest distortion of the large
scale distribution. Turbulent magnetic fields are expected to add some 
small scale distortion.

Besides the distortion of the anisotropies of the flux reaching the 
halo due to the magnetic deflections, we describe here also another effect
that can lead to anisotropies at the Earth, even if the flux were
isotropic outside the halo in the observer's reference frame. 
This is due to the fact that because of the galactic rotation there
is an electric component of the galactic field in the reference frame
in which the solar system is at rest. Then, CRs suffer a small
change in their momentum as they travel through the Galaxy. This acceleration 
is a function of the direction, inducing then anisotropies in the flux
observed at Earth in a given energy range.

We will analyze here the expected large scale distribution of
extragalactic cosmic rays arising from these effects,
taking into account the regular galactic magnetic field.
A simplified discussion of the expected Compton-Getting  
effect was presented in ref. \cite{KS}. 

\section{Galactic magnetic field effects on the extragalactic flux}

At energies below the GZK threshold cosmic rays can arrive from cosmological
distances and we assume that there is a large number of sources 
contributing to the CR flux. We will consider that the distribution of this
very large number of sources is isotropic in a
reference system that may be identified with the CMB rest frame, assuming
that the effects of an eventually anisotropic source distribution can be
neglected. Due to the solar system motion with respect to that frame,
we can model the cosmic ray flux as a dipolar distribution before entering 
the Galaxy halo. 

The presence of the galactic magnetic field along the CR trajectories 
have two effects on the CR intensity observed at Earth. These arise from
the non-trivial correspondence between the arrival direction at Earth and the 
incidence direction outside the halo due to the magnetic deflections and 
from a change in
magnitude of the particles momentum due to the electric force acting on
the particles in the observer frame. We discuss here their effects on the
flux of particles arriving to the Earth.
According to the Liouville theorem \cite{le33,sw35}
the phase space density  of cosmic rays is conserved along the particle 
trajectories. The differential flux of particles, i.e. the number of 
particles per unit solid angle and unit energy that cross a unit area 
per unit time, is related to the phase space distribution $f$ by 
$\Phi (\vec p,\vec r) = p^2 f (\vec p,\vec r)$. If CRs are only subject 
to magnetic forces along their trajectories, $p$ is conserved and thus also
the intensity of the flux is conserved, but due to the deflections of the 
particles in the galactic magnetic field, the flux corresponding to a given 
direction of incidence at the halo will appear at a different direction (or
different directions if there are multiple images) as seen from the
Earth. On the other hand, due to the high conductivity of the galactic 
medium, we can consider that there are only magnetic fields present, 
with no electric fields, in the reference frame moving locally with the plasma. 
The plasma is known to move approximately with the rotation curve of the Galaxy.
Thus, to an observer moving with the solar system, CRs traveling through 
far away regions of the Galaxy will also experience an electric force coming 
from the relative motion of the medium in which the field is purely-magnetic, 
given by
$q \vec E = (q/c) \Gamma_{\Delta V} (\vec V - \vec V_\odot) \times \vec B'$, with
$\Gamma_{\Delta V} \simeq 1$ being the Lorentz factor associated to the relative
velocity of the local system and the observer, $\Delta\vec V \equiv 
\vec V - \vec V_\odot$, and
$\vec B'$ is the magnetic field in the local frame. The effect on the 
deflection of the trajectories due to the induced electric field 
is negligible as compared to the deflection due to the magnetic field, 
as the relative velocities 
involved $\vec V - \vec V_\odot$ are at maximum only few hundred km/s, 
much smaller 
than $v$, the velocity of the ultrarelativistic CRs ($\vec V - \vec V_\odot
\ll v \simeq c$). The equation of motion for particles of charge $q$ 
in the observer's (solar system) frame can be written as
\begin{equation}
\dot{\vec p} \simeq \frac{q}{c} \big((\vec V - \vec V_\odot) \times \vec B + 
\vec v \times \vec B \big),
\end{equation}
where we can consider that $\vec B \simeq \vec B'$ as $\Gamma_{\Delta V} 
\simeq 1$.
The evolution of the direction $\hat u$ of the particle 
($\vec p = m \Gamma_v v \hat u$) is then determined by the second term,
$\hat u_f \simeq \hat u_i +(q/E) \int \vec {dl} \times \vec B$. However, the 
change in the absolute value of $\vec p$ along the trajectory is determined 
by the first term:
\begin{equation}
p_f  = p_i + \frac{q}{c^2} \int \vec dl \cdot \big((\vec V - \vec V_\odot) 
\times \vec B \big). 
\label{deltap} 
\end{equation}
Using the invariance of the phase space distribution  along the trajectories
(Liouville theorem)
we can relate the values at the solar system $f(p_0,\hat u_0)$ to the 
value outside the halo $f(p_h,\hat u_h)$
\begin{eqnarray}
f(p_0,\hat u_0) &=&  f(p_h,\hat u_h)= f(p_0,\hat u_h) +\left.\frac{\partial f}
{\partial p}\right|_{p_0} (p_h - p_0)\\
&=& f(p_0,\hat u_h) \left(1+\left.\frac{\partial \ln f}
{\partial \ln p}\right|_{p_0} \frac{p_h - p_0}{p_0} \right).
\end{eqnarray}
Using that $\Phi(p) = f(p) p^2$ we can write $\partial \ln f/\partial \ln p =
-(\gamma+2)$, where $\gamma$ is the spectral index ($\Phi (p) \propto 
p^{-\gamma}$). Then,
\begin{equation}
f(p_0,\hat u_0)=f(p_0,\hat u_h) \left(1-(\gamma+2)\frac{p_h - p_0}{p_0} \right).
\end{equation}

The phase space distribution outside the halo $f(p_0,\hat u_h)$ depends 
on the motion of the solar system with respect to the rest frame where
CRs are isotropic, with phase space distribution given by $f'(p_0)$. 
If this coincides with the CMB frame, the relation corresponds to a 
Compton-Getting dipole with an amplitude 
$\Delta=(\gamma+2) V_{CMB}/c$ \cite{KS},
\begin{equation}
f(p_0,\hat u_h)= f'(p_0) \left(1+ (\gamma + 2) \frac{\hat u_h \cdot \vec 
V_{CMB}}{c}
\right).
\end{equation}
Finally, we can relate the flux arriving to the Earth, $\Phi (p_0,\hat u_0)$,
with that outside the halo in the frame where CRs are isotropic, 
$\Phi'(p_0)$, through
\begin{equation}
\Phi (p_0,\hat u_0) \simeq \Phi'(p_0)\left(1-(\gamma+2)\frac{p_h - p_0}{p_0} 
+ (\gamma + 2)\frac{\hat u_h \cdot \vec V_{CMB}}{c} \right).
\label{phiearth}
\end{equation}
The second term in the parenthesis corresponds to the effect of the momentum 
change along the trajectory, that has to be evaluated through the integral 
in eq.~(\ref{deltap}).

\section{Quantitative results in a BSS-S model}

Both the mapping between the arrival directions at the halo and the Earth
and the change in the momentum of CRs along the trajectory 
depend on the galactic magnetic field structure and amplitude, that 
are not well known. As an illustration of the expected effects
we will present the results for a regular component of the galactic magnetic 
field modeled with a bisymmetric field with even symmetry
(BSS-S) with structure and strength very similar to those used in ref. 
\cite{stanev} but smoothed out as described in ref. \cite{toes}. 
In this model the galactic magnetic field reverses its sign between the spiral
arms of the Galaxy and the field is symmetric with respect to the Galaxy's 
mid-plane. The local value of the field is taken as
$2\ \mu$G. For the dependence on 
$z$ a contribution coming from the galactic disk and another one from the halo 
are considered:
\begin{equation}
\vec B_{reg}(x,y,z)=\vec B_{reg}(x,y,z=0) \left(\frac{1}{2\cosh(z/z_1)}
+\frac{1}{2\cosh(z/z_2)}\right)
\end{equation}
with $z_1$= 0.3 kpc and $z_2$= 4 kpc. 
We will briefly discuss how the results are modified when some other models are
considered. The effects due to the propagation in the galactic magnetic fields
on cosmic rays coming from point sources, such as the
(de)magnification of the flux, formation of multiple images, or the
modification of the spectrum measured at the
Earth, have been discussed in refs. \cite{toes,HMR00}. 

\begin{figure}
\centerline{\epsfig{file=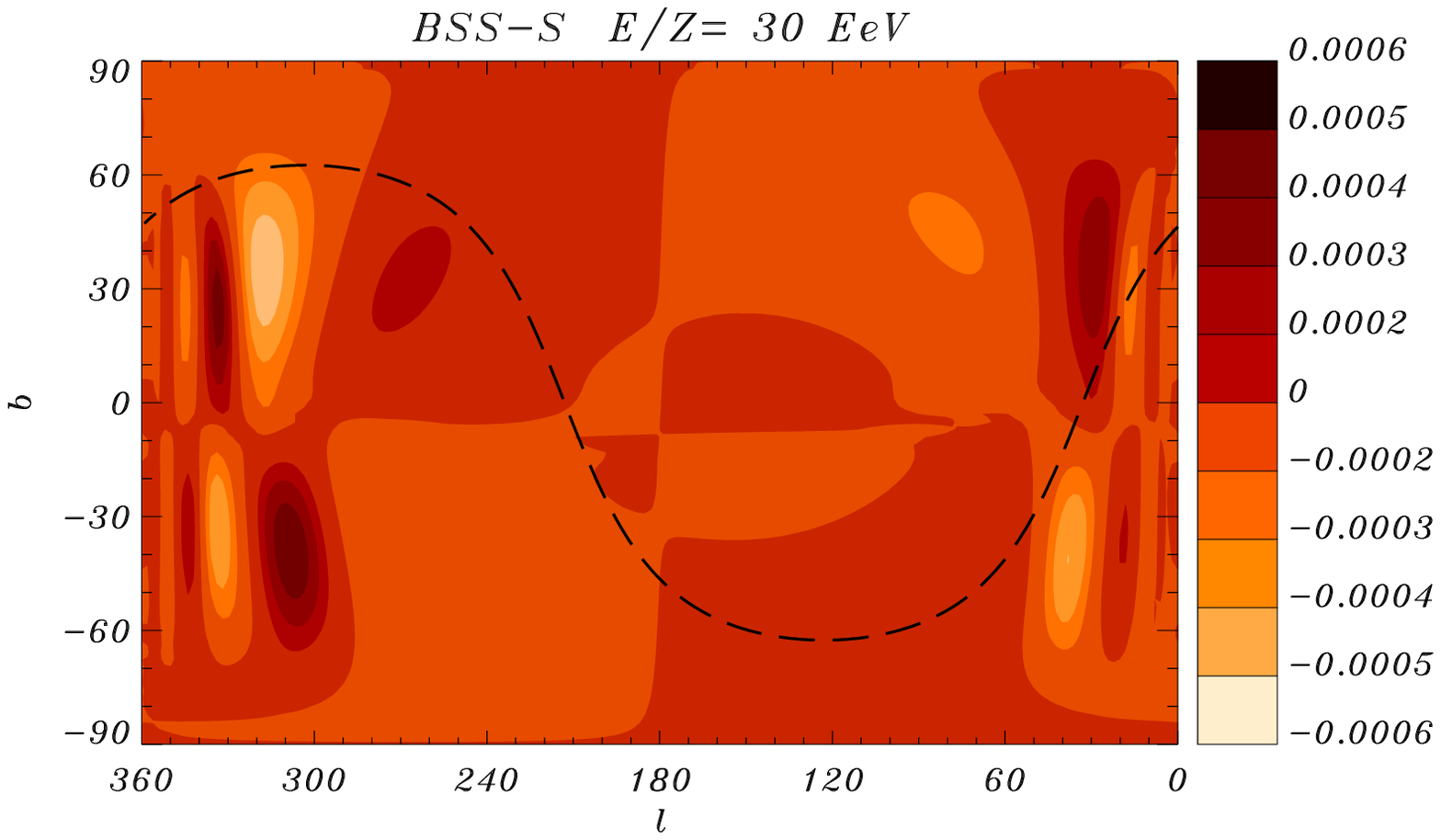,width=8cm}
\epsfig{file=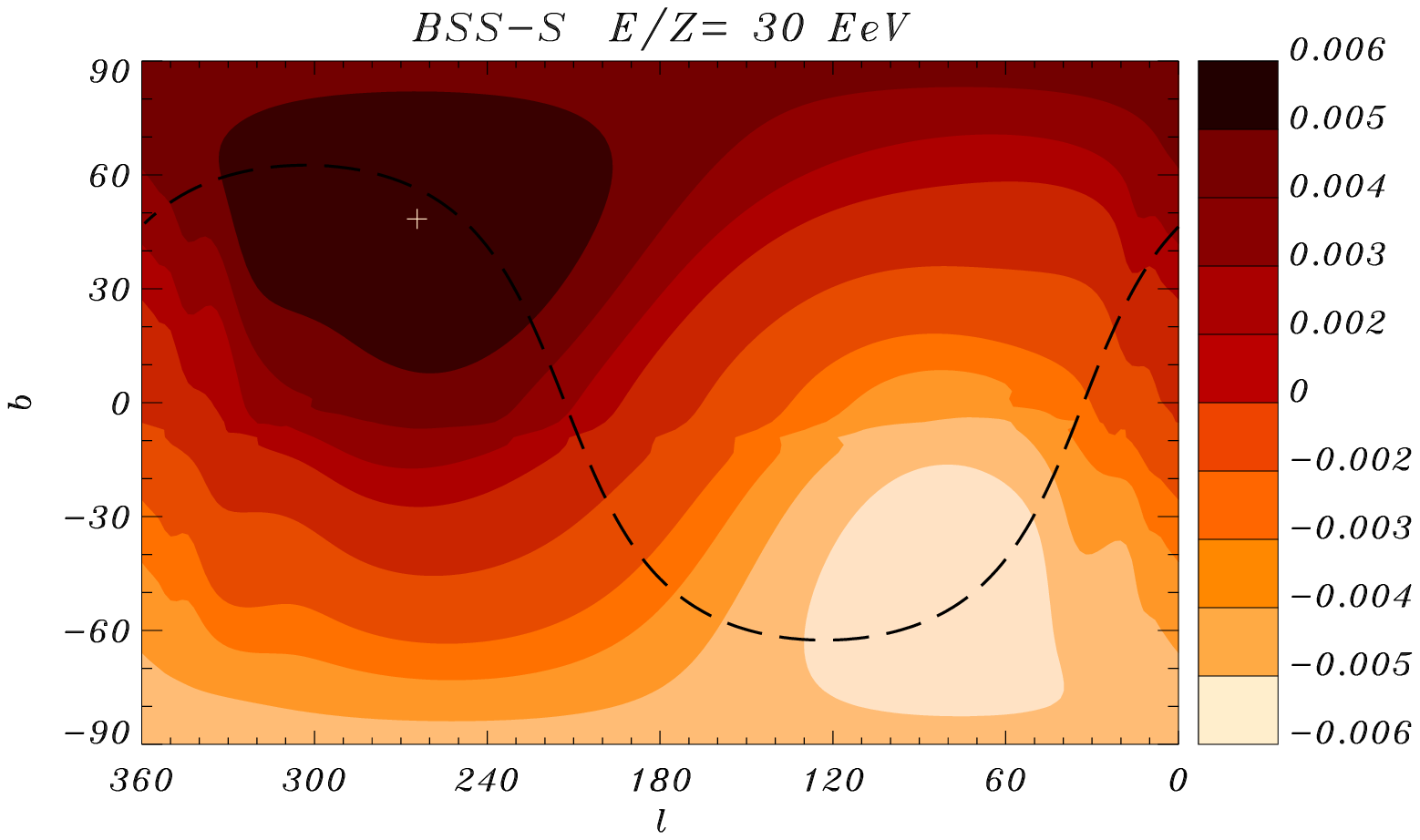,width=8cm}}
\centerline{\epsfig{file=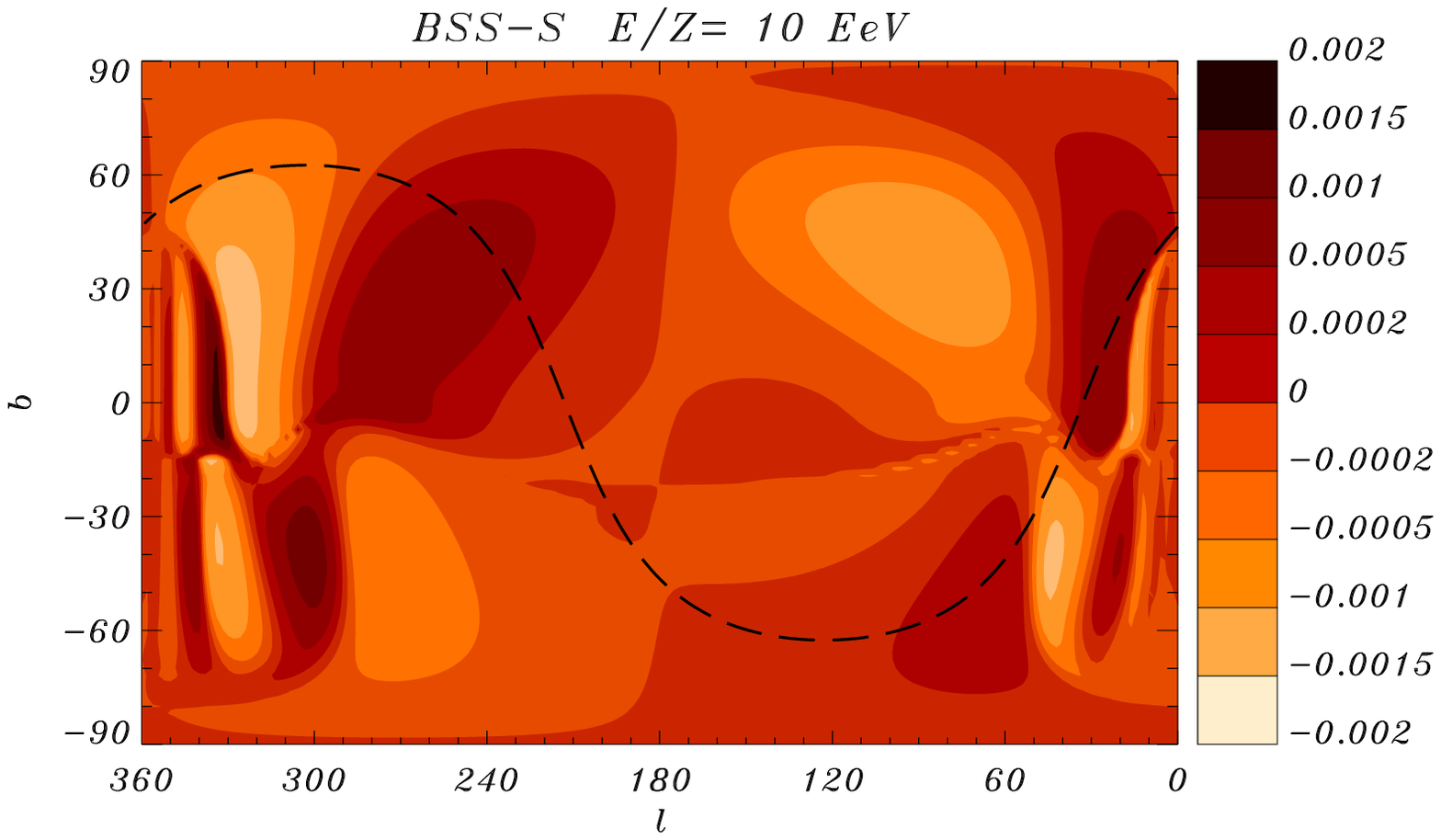,width=8cm}
\epsfig{file=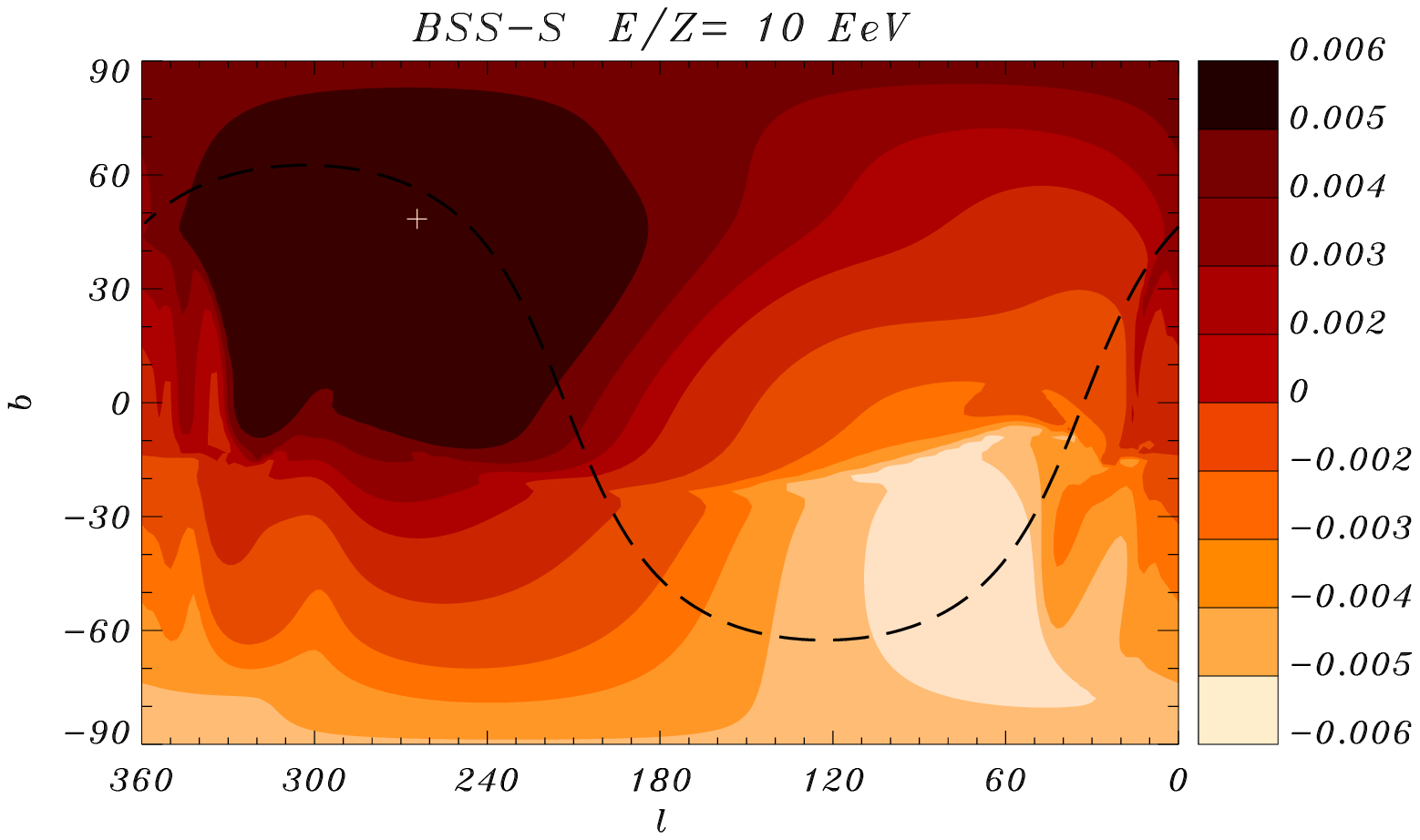,width=8cm}}
\centerline{\epsfig{file=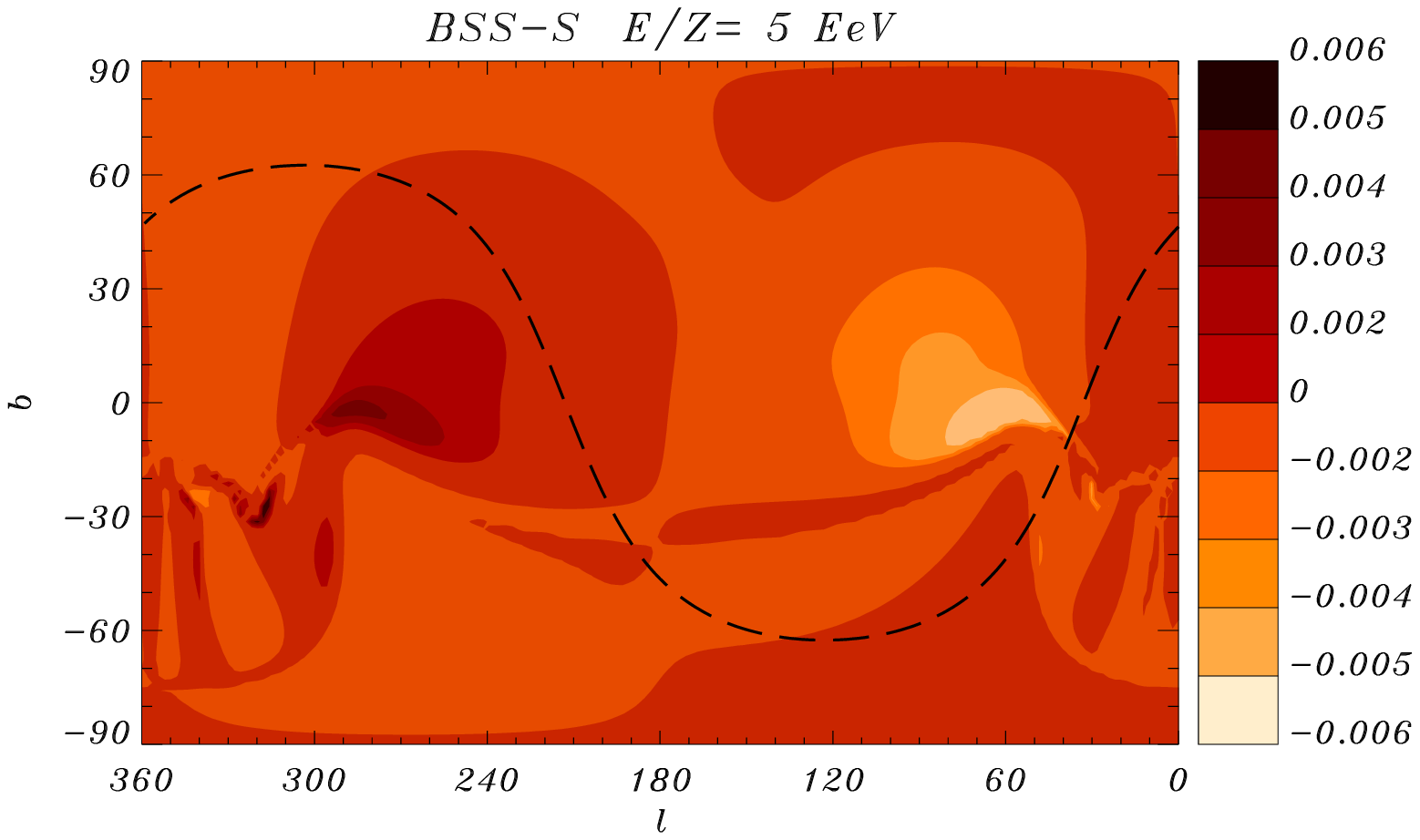,width=8cm}
\epsfig{file=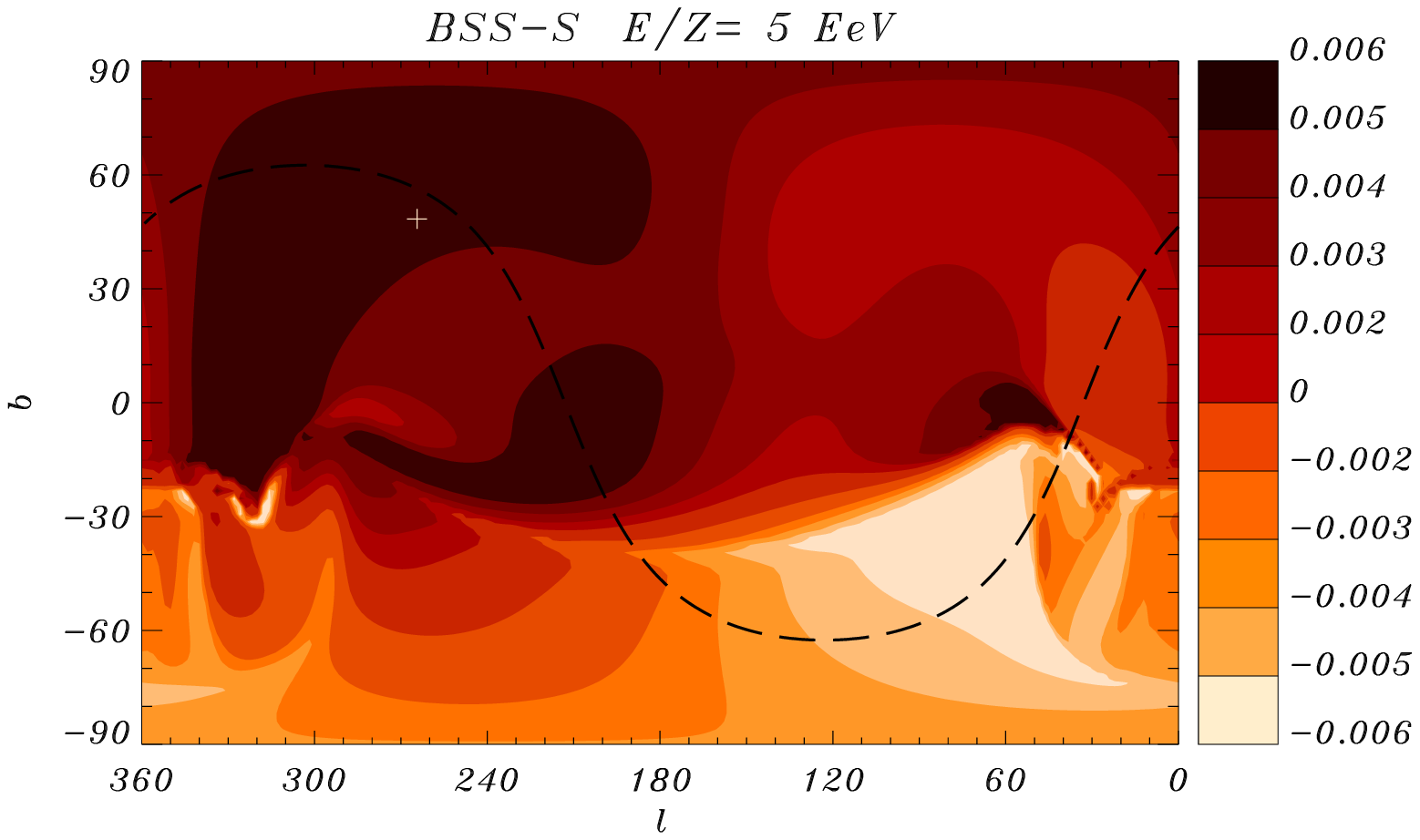,width=8cm}}
\centerline{\epsfig{file=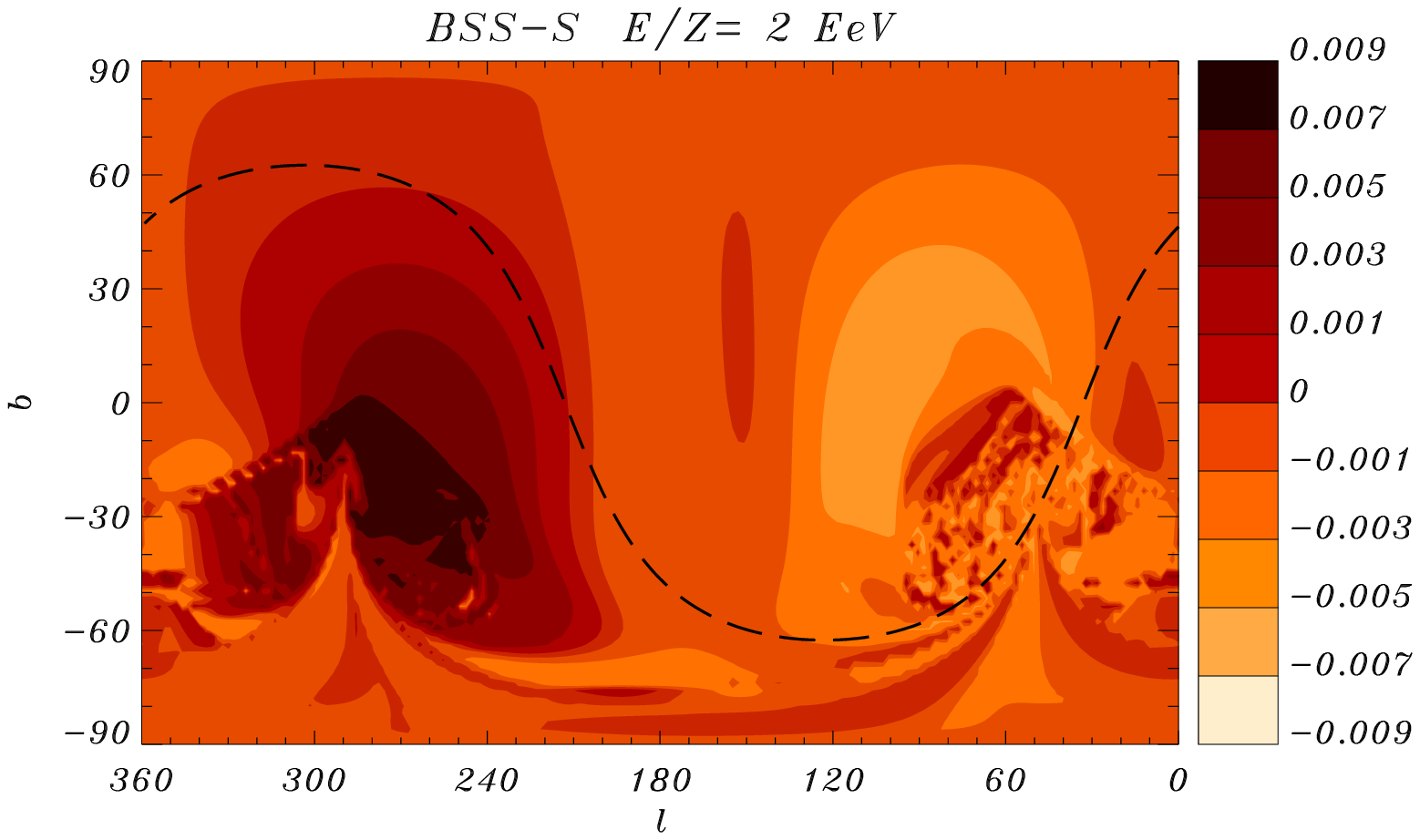,width=8cm}
\epsfig{file=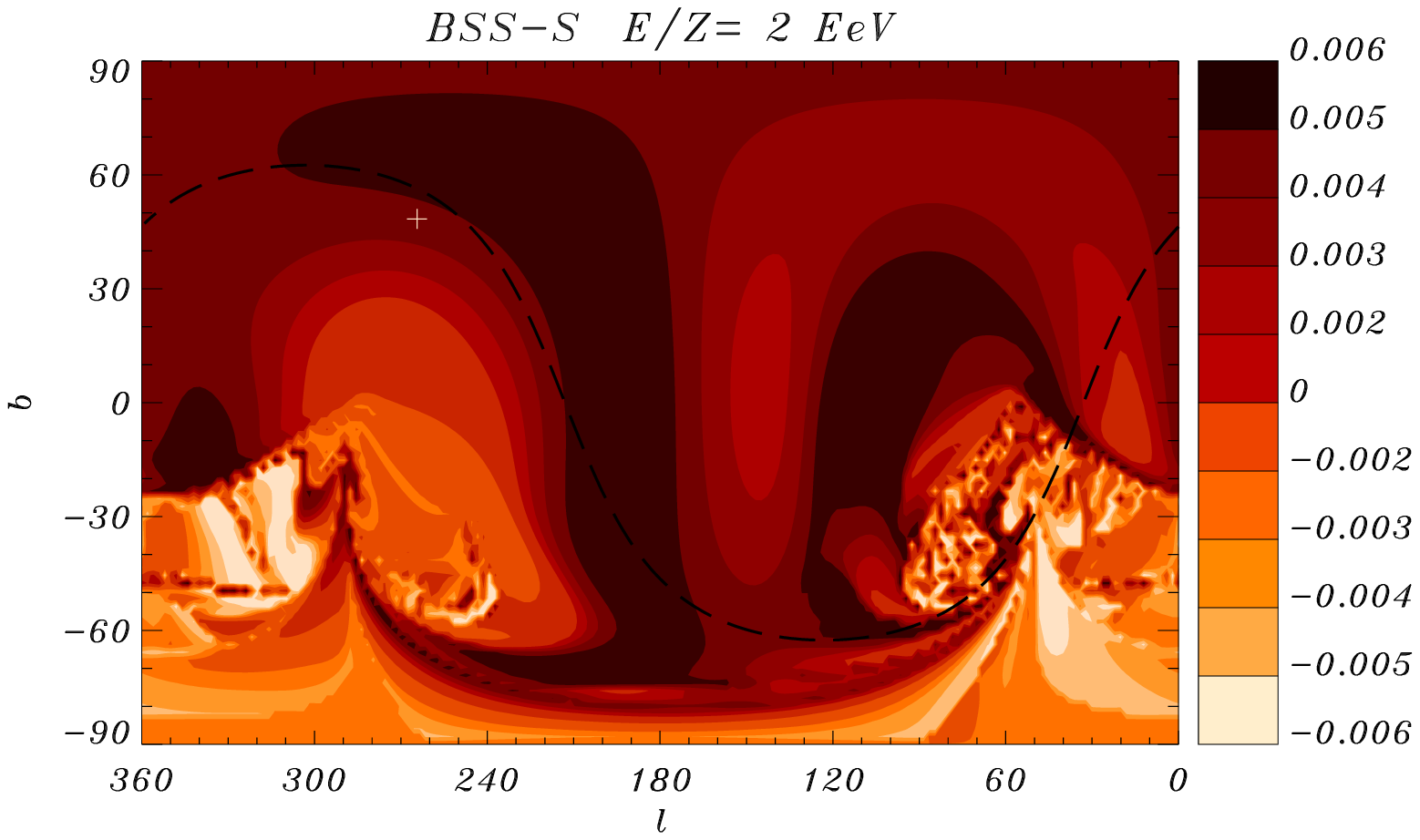,width=8cm}}
\label{fig1}
\caption{Left panels: modulation of the extragalactic flux caused by the
momentum change of CRs due to the electric force along the trajectories
for different rigidities. Right panels: Compton-Getting anisotropy 
distorted by the magnetic deflections for different rigidities.}
\end{figure}

We analyze here the effects of the galactic magnetic fields 
on a smooth flux distribution on the sky. We first consider the effect
of the the momentum change, that is a function of the rotation velocities of
the Galaxy, and then the distortion of the Compton-Getting anisotropy due to the
magnetic deflection, that is function of the solar system velocity with respect 
to the CRs rest frame. The trajectories of CRs linking the arrival direction 
at Earth to the direction of incidence at the halo are obtained 
by backtracking antiparticles leaving the Earth with a velocity opposite 
to that of the arriving CR. Performing then the integral of eq.~(\ref{deltap}) 
along the trajectories, the change of the momentum as a function of the
arrival direction is obtained. We consider that the rest frame of the plasma,
in which there is only a magnetic field, moves everywhere with a velocity
following the rotation curve of the Galaxy, that we approximate as being flat
as a function of the distance to the galactic center and with a magnitude of
220 km/s, in a clockwise direction as seen from the north galactic pole. 

We plot in the left panels in Figure 1 the resulting 
modulation factor on the halo flux considering a spectral index 
$\gamma = 2.7$ and for different values of the CR rigidities $E/Z$ 
(corresponding to the second term in the right hand side of 
eq.~(\ref{phiearth})). The galactic magnetic field model considered has 
only azimuthal and radial components (no component in the direction 
perpendicular to the galactic plane). As the relative velocities
$\vec V  - \vec V_\odot$ have also only azimuthal and radial components,
the electric force is always perpendicular to the galactic plane, pointing
up or down in different spiral arms due to the reversals of the magnetic field.
For high rigidities, the trajectories are almost straight, and then for 
particles traveling in the galactic plane direction the effect is negligible
as the electric force is orthogonal to the trajectory and thus does not change
the magnitude of the momentum. 
Also for $l=0^\circ$ and $l=180^\circ$ the effect 
is negligible as for those directions the force is very small because the 
relative velocities and the magnetic field are almost parallel.\footnote{
These features would change if there is for instance a dipole
contribution to the magnetic field, leading to a vertical non-vanishing 
component, which could be significant in the central regions of the Galaxy.} 
The effect has 
a clear asymmetry with respect to the galactic plane with regions showing
an excess of flux in the northern hemisphere accompanied by a region with a
flux deficit in the southern hemisphere at the same galactic longitude, and
vice-versa. This is due to the fact that an upward pointing electric field 
at a given longitude accelerate particles arriving from the south, while 
decelerates particles arriving from the north. For smaller rigidities
this pattern becomes distorted because the trajectories are increasingly
deflected in the magnetic field. On the other hand, the trajectories inside 
the halo are longer and the magnitude of the momentum change is larger,
leading to an increase of the modulation of the flux, as is evidenced in
the change in the color scale of the figures for decreasing rigidities.

The right panels in Figure 1 show the modulation factor on the halo 
flux due to the Compton-Getting effect if the rest frame where CRs are 
isotropic coincides with the CMB one, considering a spectral index 
$\gamma = 2.7$ and for different values of the CR rigidities  
(corresponding to the third term in the right hand side of 
eq.~(\ref{phiearth})). The direction of the original dipolar
component of the extragalactic cosmic ray flux  points to the 
$(b,l) = (48.4^{\circ},264.4^{\circ})$ direction, indicated by the white
cross in the plots and its amplitude is $\Delta\simeq 0.6$\%. 
As the rigidity decreases the deflections of the particles grow and
the distortions of the dipolar modulation become larger and it is apparent 
that higher order harmonics become important.

The combined effect of the variation of the momentum along the trajectory
 and the magnetic
deflection distortion of the Compton-Getting dipole is shown in Figure
2 for the same rigidities as in Figure 1.
\begin{figure}
\centerline{\epsfig{file=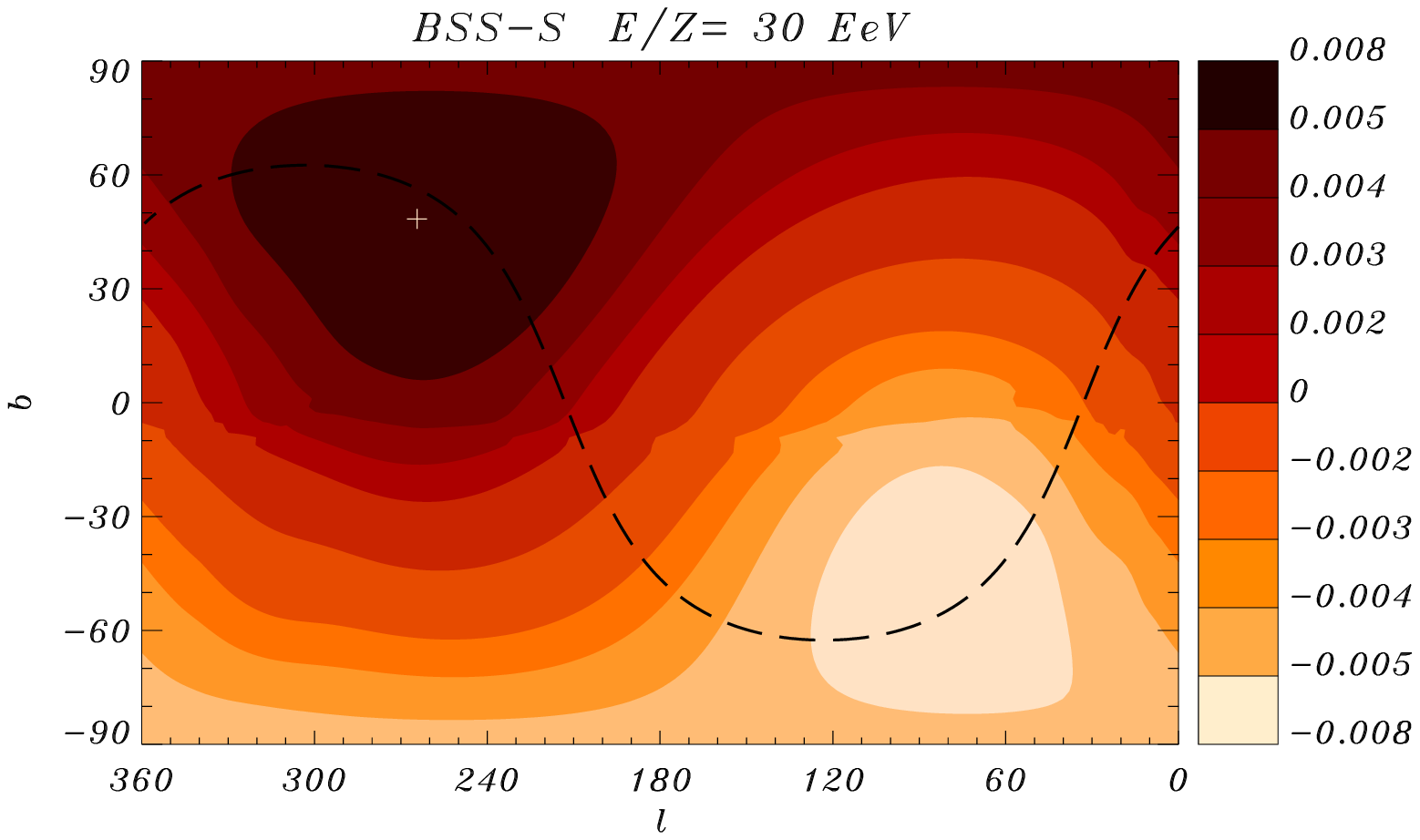,width=8cm}
\epsfig{file=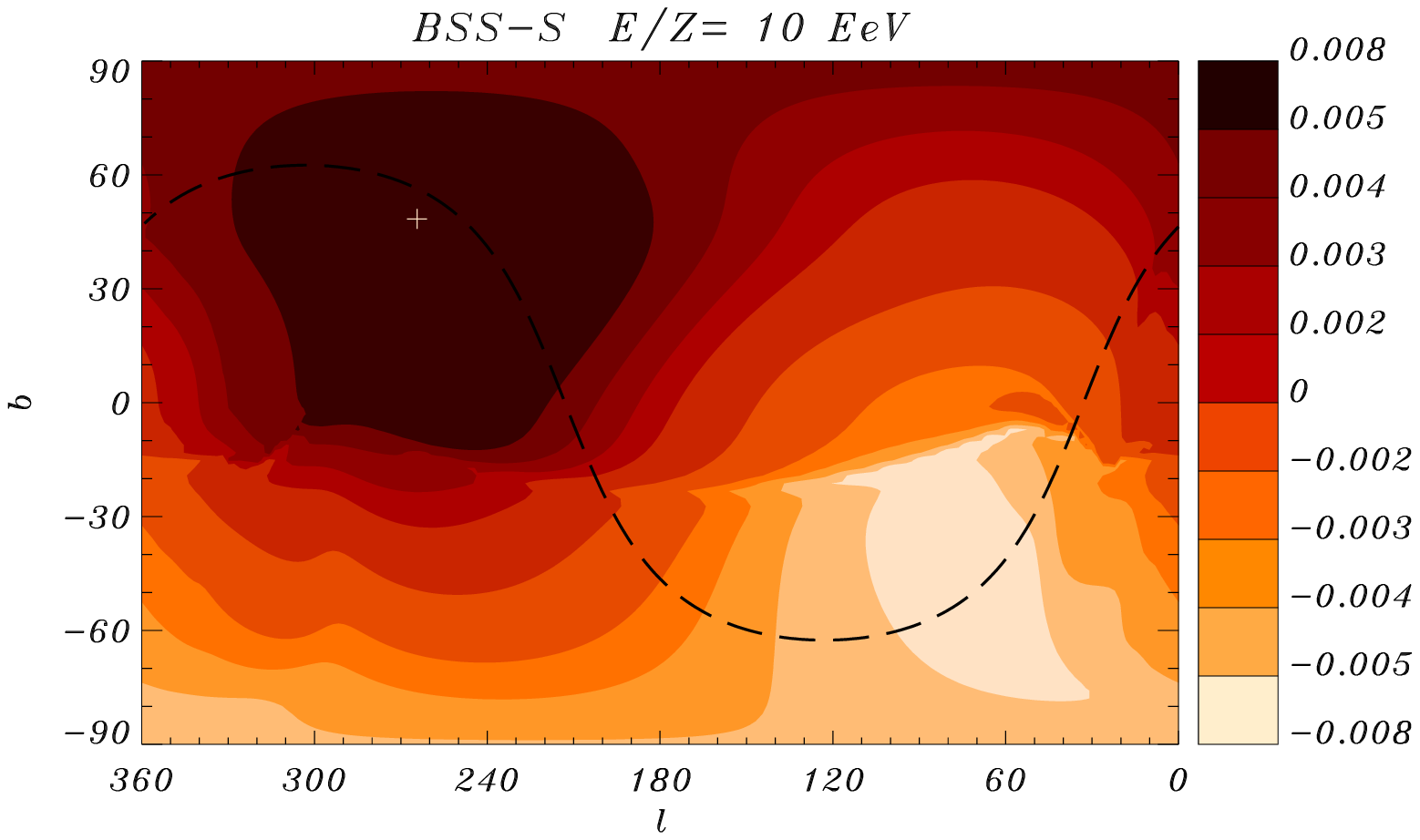,width=8cm}}
\centerline{\epsfig{file=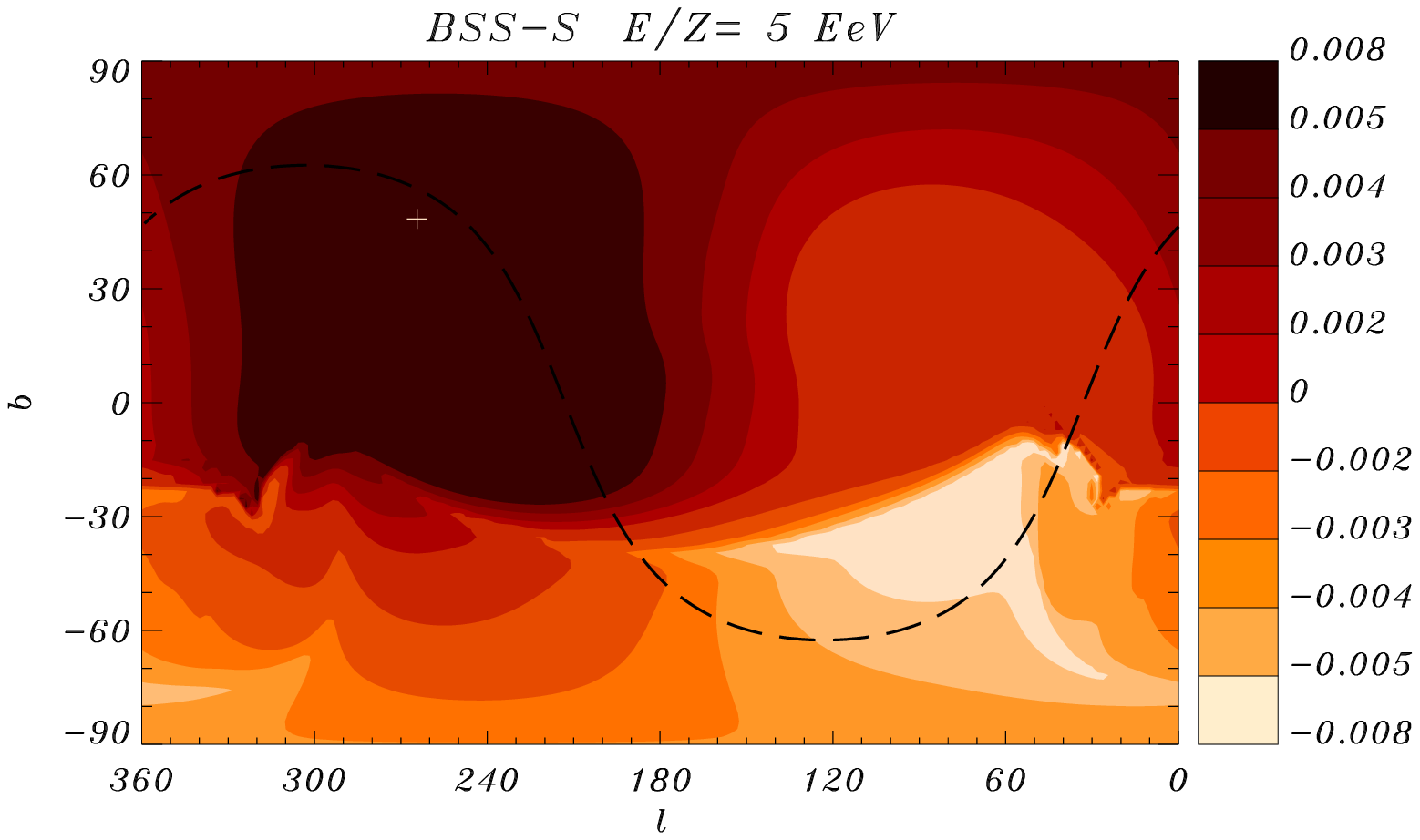,width=8cm}
\epsfig{file=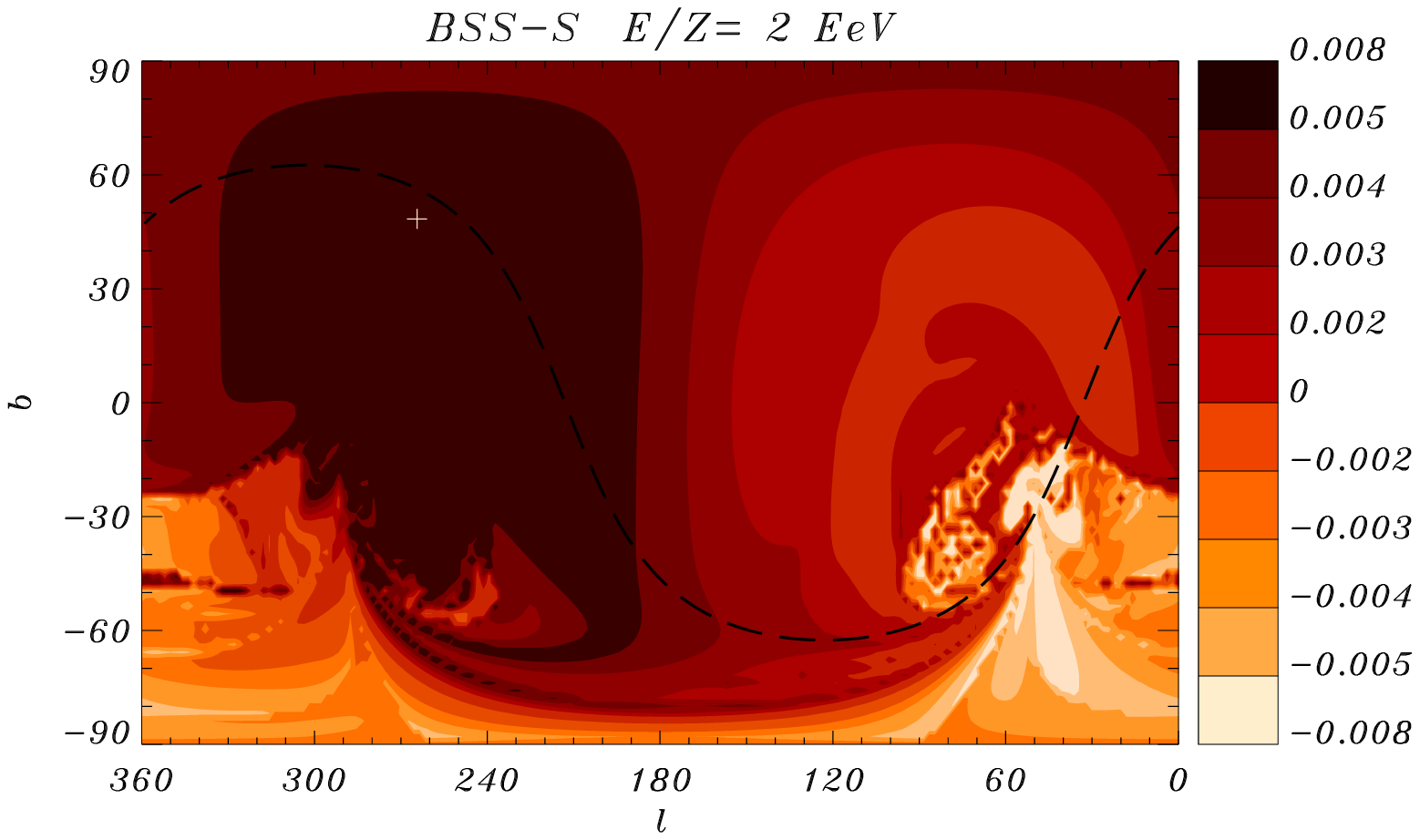,width=8cm}}
\label{fig2}
\caption{Total modulation of the flux for different rigidities.}
\end{figure}

Figure 3 shows the amplitude of the dipolar component arising from each
effect, as well of the total one.
For the particular orientation of the 
dipole considered (CMB rest frame) and the galactic magnetic field
model considered, the dipolar amplitude of the Compton-Getting 
anisotropy is increasingly suppressed
as the rigidity decreases below 10 EeV, 
reaching a $50\%$ suppression for rigidities
around $2-3$ EeV. On the other hand, the anisotropy arising from
the momentum variation along the trajectories grows for energies below 
10 EeV and around 2 - 3 EeV its dipolar amplitude is similar to the 
Compton-Getting one.
\begin{figure}
\epsfig{file=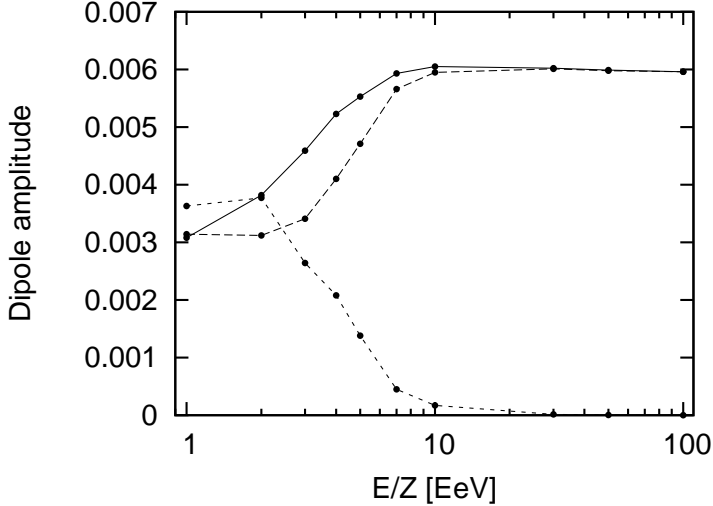,angle=-90,width=10cm}
\caption{Amplitude of the dipolar component of the flux modulation 
as a function of the rigidity for the Compton-Getting anisotropy
distorted by the magnetic deflections (long-dashed line), for the momentum
change induced anisotropy (short-dashed line) and for the sum of both
effects (solid line).}
\end{figure}
A curious fact that is evident in the two lower energy panels 
($E/Z = 5$ EeV and 2 EeV) of Figure 1 is that the anisotropy generated 
through the momentum variation along the trajectories (left panels) have
flux excesses (and deficits) located roughly in the same directions
where the deficits (and excesses) corresponding to the Compton-Getting
anisotropy are located (right panel). As a result of this the total combined
anisotropy shown in Figure 2 has a more dipolar-like shape at these lower 
energies than any of the two individual effects, and the total dipolar
amplitude is larger as shown in Figure 3. This is curious because the
amplitude of both effects depend on different unrelated velocities.
For example, had the Milky Way be moving in the opposite direction 
with respect to the CMB rest frame, the Compton-Getting modulation of 
the flux displayed in the right panels in Figure 1 would have had the
opposite sign, while the left panels would have been unchanged, and the
compensation would not have occurred.
These results depend thus on which is the rest frame where CRs are isotropic.

At rigidities below 10 EeV the contribution from higher order 
harmonics becomes significant as it can be qualitatively appreciated in
Figure 1. By expanding the flux reaching the Earth 
as a series of spherical harmonics with coefficients $a_{\ell m}$,
\footnote{we use the real spherical harmonics base, so that the $a_{\ell m}$
are real.}
\begin{equation}
\Phi(\hat u)=\sum^{\infty}_{l=0}\ \sum^{\ell}_{m=-\ell}\
a_{\ell m}\ Y_{\ell m}(\hat u),
\end{equation}
we can quantify the relative amplitude of the different multipoles
through the angular power spectrum $C_\ell$, that is given by the average 
over the possible $m$ values of the $a_{\ell m}^2$ coefficients,
\begin{equation}
C_\ell = \frac{1}{2\ell+1}\ \sum_{m=-\ell}^{\ell}\ a_{\ell m}^2.
\end{equation}
We show in Figure 4 the normalized contribution for the 
three lowest order terms, the dipole, the quadrupole and the octupole
as a function of the rigidity.
\begin{figure}
\epsfig{file=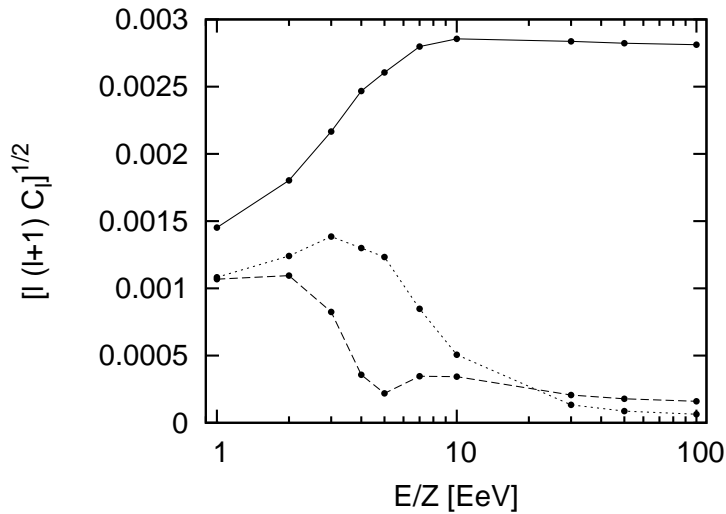,angle=-90,width=10cm}
\label{fig3}
\caption{Lowest order power spectrum terms as a function of the rigidity:
dipole ($\ell = 1$, solid line), quadrupole ($\ell = 2$, short dashed line)
and octupole ($\ell = 3$, long dashed line).}
\end{figure}
At the highest rigidities only the dipole component is significant, 
but as the rigidity decreases the quadrupole and octupole components
become important.

The previous results have all been obtained for the BSS-S 
galactic magnetic field model. The main results still hold
qualitatively for  an ASS-A model (i.e. with no reversals and
antisymmetric with respect to the Galactic plane) with the same local
amplitude and halo parameters.

The results are relevant for energies such that the extragalactic
component of cosmic rays is dominant, that is above $\sim$ 4 EeV if
the ankle marks the galactic to extragalactic transition or at even
lower energies in the $e^{+}e^{-}$ pair production dip scenario for
the ankle. On the other hand, we have assumed that the flux of
extragalactic cosmic rays can be approximated as a diffuse dipolar
like flux. This approximation holds only for a very large number of
sources distributed at cosmological distances. Thus, the energy should
be small enough so that the GZK effect does not considerably attenuate
the flux from distant sources. For energies below 30 EeV cosmic rays 
can arrive from distances of about 1 Gpc, then this is a safe upper energy
limit to consider. 

If extragalactic protons are the dominant component of the flux at the
ankle energies and below, as in the $e^{+}e^{-}$ pair production dip  
scenario for the ankle, for large enough intergalactic magnetic field
values the extragalactic flux may experience a diffusive motion at the
lower energies. It has been pointed out that this could lead to the 
appearance of a magnetic horizon: the time needed for cosmic rays to
arrive at Earth from distances larger than this horizon is typically
larger than the age of the universe, suppressing the flux arriving from
them \cite{maghor}. This affects the anisotropy of cosmic rays
arriving to the halo of the Galaxy as discussed in ref. \cite{al08}.
If the particles with energies $10^{17} - 10^{18}$ eV arrive
predominantly from the closest source through a diffusive propagation, 
the flux reaching the galactic
halo will be dipolar-like with amplitude about $10^{-2} - 10^{-3}$. In
this case, the propagation of cosmic rays in the galactic magnetic field 
will also affect the anisotropies observed at Earth, and an analysis along 
the lines of this work still holds,
although the direction of the dipole would not coincide with the CMB
one, but would point to the direction to the closest source.

\section{Conclusions}

We have analyzed in detail the impact of the propagation through the
galactic magnetic field on the large scale anisotropies of extragalactic 
cosmic rays. As the propagation is a function of the cosmic ray rigidity, 
the resulting anisotropy is also a function of the
rigidity. On one hand, the Compton-Getting dipolar anisotropy 
resulting from the motion of the observer with respect to the frame 
in which the cosmic rays are isotropic is distorted by the magnetic 
deflection of the trajectories in the galactic magnetic field. 
In the absence of magnetic field effects, the expected dipole amplitude 
is of about $0.6\%$ if the CRs rest frame coincides with the CMB one. 
The distortion due to the magnetic 
deflections modifies this amplitude, leading to a suppression by a factor of 
up to $50\%$  at energies below 3 EeV 
for the BSS-S galactic magnetic field considered.
Moreover, as a consequence of the propagation the general pattern of the
flux is distorted and significant higher order harmonics appear in the
expected distribution of arrival directions.

On the 
other hand, even if the observer's rest frame is not moving with respect to the
CRs frame and the flux of CRs entering the halo were perfectly isotropic, 
the flux distribution observed at Earth would be
affected by the galactic magnetic field. This is due to the fact that the 
motion of the galactic medium (and hence of the magnetic field) following the 
galactic rotation leads to an electric field in the 
reference frame moving with the solar system. Then, CRs suffer a small
change in their momentum as they travel through the Galaxy. This acceleration 
is a function of the direction, inducing then anisotropies in the flux
observed at Earth in a given energy range. This effect grows with decreasing 
energy, reaching $\sim 0.3\%$ at energies below 2 - 3 EeV.

Let us mention that in this work we considered the cosmic ray distribution 
in the whole sky, but for specific experiments on Earth only part
of the sky is observed. The details of the results obtained in a given
cosmic ray observatory will then depend on the declination dependence of
its exposure and hence on the site location. Anyhow the results will be 
qualitatively similar to those obtained for the full sky.
The detection of this effect is a challenge for present experiments.
The Pierre Auger Observatory has recently published upper limits 
for the first harmonic amplitude in this energy range. The $99\%$ CL upper
limit on the amplitude of the equatorial component of a dipole is 
about $2\%$ at 1 EeV and about $10\%$ at 10 EeV \cite{au09}.
If at EeV energies cosmic rays are predominantly extragalactic,
with a few fold increase in the present statistics this experiment would become
sensitive to a dipole anisotropy of the amplitude discussed here. 

\acknowledgments

This work is supported by ANPCyT (grant PICT2006 13334) and CONICET (grant PIP
01830).

\end{document}